\documentstyle[12pt]{article}

\begin{document}

\baselineskip=7mm
\renewcommand{\arraystretch}{1.3}

\newcommand{\cf}{{ f}}
\newcommand{\TeV}{\,{\rm TeV}}
\newcommand{\GeV}{\,{\rm GeV}}
\newcommand{\MeV}{\,{\rm MeV}}
\newcommand{\keV}{\,{\rm keV}}
\newcommand{\eV}{\,{\rm eV}}
\newcommand{\Tr}{{\rm Tr}\!}
\newcommand{\be}{\begin{equation}}
\newcommand{\ee}{\end{equation}}
\newcommand{\bea}{\begin{eqnarray}}
\newcommand{\eea}{\end{eqnarray}}
\newcommand{\ba}{\begin{array}}
\newcommand{\ea}{\end{array}}
\newcommand{\bmat}{\left(\ba}
\newcommand{\emat}{\ea\right)}
\newcommand{\refs}[1]{(\ref{#1})}
\newcommand{\ler}{\stackrel{\scriptstyle <}{\scriptstyle\sim}}
\newcommand{\ger}{\stackrel{\scriptstyle >}{\scriptstyle\sim}}
\newcommand{\lag}{\langle}
\newcommand{\rag}{\rangle}
\newcommand{\ns}{\normalsize}

\begin{titlepage}
\title{{\Large\bf Cosmological Implications  of 
Radiatively Generated  Axion Scale}\\
                          \vspace{-4.5cm}
                          \hfill{\ns KAIST-TH 12/96\\}
                          \hfill{\ns CBNU-TH 960707\\}
                          \hfill{\ns SNUTP 96-074\\}
                          \hfill{\ns hep-ph/9608222\\}
                          \vspace{3.5cm} }

\author{Kiwoon Choi\\[.5cm]
  {\ns\it Department of Physics, Korea Advanced Institute of Science
           and Technology}\\
  {\ns\it Taejon 305-701,  Korea} \\[.5cm]
  {Eung Jin Chun}\\[.5cm]
  {\ns\it Department of Physics, Chungbuk National University}\\
  {\ns\it Cheongju, Chungbuk 360-763,  Korea} \\[.5cm]
  {Jihn E. Kim }\\[.5cm]
  {\ns\it Department of Physics and Center for Theoretical Physics}\\
  {\ns\it Seoul National University, Seoul 151-742, Korea}\\
  }
\date{}
\maketitle
\begin{abstract} 
\baselineskip=7.2mm  {\ns}
We study cosmological implications of supersymmetric axion models
in which the axion scale is generated radiatively.
Such models lead to the so-called thermal inflation  and  
subsequent reheating should be constrained
not to yield  a too large
axion energy density at the time of nucleosynthesis.
We examine how plausible it is that   this nucleosynthesis constraint
is satisfied for   both hadronic and 
Dine-Fischler-Srednicki-Zhitnitskii type axion models.
Baryogenesis and the possibility 
for raising up the cosmological upper bound on the axion scale in 
thermal inflation scenario are also discussed.
\end{abstract}

\thispagestyle{empty}
\end{titlepage}

One of the attractive solutions to the strong
CP problem is to introduce 
an anomalous Peccei-Quinn (PQ) symmetry $U(1)_{PQ}$ \cite{pq}. 
This solution predicts a pseudo-Goldstone boson, 
the (invisible) axion \cite{k,dfs}, 
whose decay constant $F_a$ is tightly constrained
by astrophysical and cosmological arguments.
The allowed band of the axion scale $F_a$  lies 
between $10^{10}$ GeV and $10^{12}$ GeV \cite{kim}
which is far away from the already known two mass scales, 
the electroweak scale and the Planck scale $M_P=1/\sqrt{8\pi G_N}$.
It is certainly desirable that this intermediate scale
appears  as a dynamical  consequence when the known
mass scales are set up in the theory.
This indeed happens \cite{msy} in some class of 
spontaneously broken supergravity models which are commonly considered
as the underlying structure of the supersymmetric standard model.
Such models typically contain two basic mass scales, $M_P$ and the scale of
local supersymmetry breaking $M_S$ in the hidden sector leading to
$m_{3/2}=M_S^2/M_P=10^{2}\sim 10^3$ GeV. Supergravity interactions 
then generate  soft supersymmetry breaking terms 
in the supersymmetric standard model sector which are
of order
$m_{3/2}$. In this scenario,
radiative corrections to the Higgs doublet mass-squared
associated with the large top quark Yukawa coupling  
can naturally lead
to the electroweak symmetry breaking at the scale $M_W\simeq m_{3/2}$.
When the PQ fields which are responsible for the spontaneous
violation of $U(1)_{PQ}$ correspond to flat directions of the model,
the intermediate axion scale $F_a$  can  also  be radiatively
generated  in terms of  $M_P$ and $m_{3/2}$. 
In such a scheme, as was recently emphasized, the early universe
experiences  the so-called thermal inflation
and subsequently a period 
dominated by  coherently oscillating flaton fields \cite{ls}.
The aim of this paper is to examine cosmological implications of 
PQ flatons  in supergravity models with a radiative mechanism
generating the axion scale.

One possible cosmological consequence of PQ flatons is the impact on the
big-bang nucleosynthesis  through their decay into axions.  
In the scheme under consideration,
PQ flatons have generally  order-one coupling to the Goldstone boson 
(the axion) in the unit of $1/F_a$ \cite{cl}. 
As we will argue later,
axions produced by decaying flatons are hardly thermalized.
In this paper, we first consider  the energy density of
these unthermalized axions at the time of nucleosynthesis
together with its implications
for  both
hadronic axion
models \cite{k}  and   
Dine-Fischler-Srednicki-Zhitnitskii (DFSZ) type
models \cite{dfs}.
Even when one takes a rather conservative limit
on the axion energy density,
this consideration provides a meaningful restriction
for generic hadronic axion models and also for
DFSZ type models with a rather large flaton mass.

As another cosmological implications  of PQ flatons,
we consider  the possibility 
of raising up the cosmological upper bound
on the axion scale $F_a$  through the late time entropy production \cite{dfst}
by oscillating PQ flatons. 
We argue that $F_a$ can be pushed up to about $10^{15}$
GeV without any cosmological difficulty in thermal
inflation scenario.
Finally,  we point out that
the Dimopoulos-Hall (DH) mechanism \cite{dh} for a late time baryogenesis
can be naturally implemented  in thermal inflation scenario.
In the conclusion, we note that the case of $n=2$ or 3 [see Eq.~(3) below]
provides a very concordant cosmological scenario.  
\bigskip

We begin by describing how the intermediate axion scale can be
radiatively generated in supergravity models in which  the PQ fields
correspond to flat directions.
Let us consider a variant of the model of Ref.~\cite{msy} with 
superpotential  
\be \label{supo}
 W = k{\phi_1^{n+2} \phi_2 \over M_P^n } + h {\phi_1^{n+1} H_1 H_2 \over
     M_P^n} + h_N N N \phi_1 + h_L L H_2 N + \cdots 
\ee
where $H_{1,2}$ are the usual Higgs doublets,
$N$ is the right-handed neutrino component and the ellipsis
denotes  the supersymmetric standard model part of the superpotential.  
In  order to implement the PQ symmetry, two gauge  singlet superfields  
$\phi_{1,2}$ with PQ charges $q_{1,2}$ are introduced.
The structure of the superpotential is determined by the PQ charge 
assignment: $q_2=-(n+2)q_1$,  $q_{_{H_1}}+q_{_{H_2}}=-(n+1)q_1$ and so on.
Obviously the PQ fields $\phi_1$ and $\phi_2$ correspond
to flat directions when nonrenormalizable interactions and
supersymmetry breaking effects are ignored.
This model can be considered as a supersymmetric generalization
of the DFSZ axion model (but endowed
with a radiative mechanism generating the axion scale) in the sense
that the Higgs doublets carry nonzero PQ charges.
Note that the second term in the superpotential
yields the correct scale for the Higgs
mass parameter $\mu =h \lag \phi_1 \rag^{n+1}/M_P^n$ upon 
spontaneous breaking of the PQ symmetry \cite{kn}.  
Taking into account the  radiative effects of 
the strong Yukawa coupling
$h_NNN\phi_1$, the  soft mass-squared of 
$\phi_1$  becomes  {\it negative} at scales around $F_a\simeq \lag \phi_1\rag$,
and thereby driving $\phi_1$ to develop vacuum expectation value at
an intermediate scale.
This Yukawa coupling is also necessary to keep the field $\phi_1$ in thermal
equilibrium at high temperature $T> m_1$ for which $\lag \phi_1 \rag =0$.
Neglecting the field $\phi_2$, the renormalization group improved
scalar potential for the singlets
is given by
\be \label{eq1}
  V = V_0 - m_1^2 |\phi_1|^2 + k^2 {|\phi_1|^{2n+4} \over M_P^{2n} }\,,
\ee
where $m_1^2$ is positive and of order $m_{3/2}^2$,
and  $V_0$ is a constant of order $m_{3/2}^2F_a^2$ which is introduced
to make $V(\lag\phi_1\rag)=0$.
Clearly the minimum of this scalar potential breaks $U(1)_{PQ}$ 
by
\be \label{Fa}
\lag \phi_1\rag \simeq F_a \simeq (m_{3/2}M_P^{n})^{1/n+1},
\ee
where we have ignored the coefficients of order unity.
The integer $n$ fixes the size of the axion scale.  For the smallest
value $n=1$, the axion scale $F_a \simeq \sqrt{m_{3/2}M_P}$  fits into 
the usual allowed band of the axion scale: 
$ 10^{10} \GeV \ler F_a \ler 10^{12} \GeV$.
Later we will argue that the upper bound on $F_a$ can be relaxed 
and thus a bigger value of $n$ is allowed also.

The above radiative 
mechanism generating  the axion scale has substantial 
influence on the history of the universe \cite{ls,ls2}.  At high temperature,
$\phi_1$  receives a 
thermal mass $\delta m_1^2\simeq  |h_N|^2T^2\gg m_1^2$ leading to 
$\lag \phi_1 \rag =0 $.
This thermal  mass is  generated by  right-handed neutrinos
in the thermal bath.
Note that the right-handed neutrino $N$
becomes massless when $\lag \phi_1\rag=0$ and thus  
copiously produced  when $T\gg m_1$.
During this period, $\lag \phi_2\rag =0$ also. When the temperature  
falls below $T \simeq V_0^{1/4}$, which is about 
$\sqrt{m_{3/2} F_a}$, the universe is dominated by
the  vacuum energy density $V_0$
and thus there appears a short period of thermal inflation.  
Below $T < m_1 \simeq m_{3/2}$,  the effective mass of $\phi_1$ becomes
negative and then  $\phi_1$ develops an intermediate scale
VEV  given by  Eq.~\refs{Fa}. With  $\langle \phi_1\rangle \simeq
F_a$, the other flaton field
$\phi_2$ develops also a VEV of order $F_a$
through  the $A$-type soft SUSY
breaking term, $kA\phi_1^{n+2}\phi_2/M_P^n$, in the scalar potential.
This procedure makes the thermal inflation end and subsequently
the early  universe experiences a period 
dominated by
coherently oscillating PQ flaton fields $\phi_1$ and $\phi_2$.
More precisely, the oscillating flaton corresponds to
a combination of the two complex scalar fields $\phi_1$ and $\phi_2$
which is orthogonal to the axion field
$a= \sum_i c_i \, {\rm arg}(\phi_i)$ where
$c_i=q_i\lag\phi_i\rag^2/F_a$.
\bigskip

{\em NS bound. }
After the period of coherent oscillation, the universe would be reheated by
the decay products of the oscillating flaton $\varphi$.
A feature peculiar to the PQ flatons is that
their decay products include axion as one of the main components 
\cite{cl,ls2}. 
The energy density of these axions
at the time of nucleosynthesis (NS), $(\rho_a)_{NS}$, should satisfy
the conventional nucleosynthesis bound on the extra
energy density:
\be
\left({\rho_a\over \rho_{\nu}}\right)_{NS}\leq \delta N_{\nu}.
\ee
Here $\rho_{\nu}$ denotes the energy density of a single species
of relativistic neutrino and $\delta N_{\nu}$ is the 
number of extra neutrino species allowed by nucleosynthesis.
In the past, $\delta N_{\nu}$ has been argued to be $0.3$ or 
even smaller as $0.04$ \cite{nsold}.
However, although claimed to be quite conservative, 
more careful recent analyses do not  exclude even 
$\delta N_{\nu}=1.5$ \cite{nsnew}.
Here we do not take any specific value of $\delta N_{\nu}$,
but examine the implications of the above NS bound for $\delta N_{\nu}
=0.1 \sim 1.5$ with the hope that one can push $\delta N_{\nu}$
down to the value $0.1$ in the future.

Before evaluating  $(\rho_a)_{NS}$,
let us first determine  the reheat temperature $T_{RH}$
by  parameterizing  the width of the flaton
decay into thermalizable particles as 
$\Gamma_{\varphi}=B_a^{-1}M_{\varphi}^3/64\pi F_a^2$.
Here $M_{\varphi}$ denotes the flaton mass
and the prefactor $B_a^{-1}$ 
will be presumed  to be of order 10, which is a proper choice
for $(\rho_a)_{NS}$ to satisfy the above NS bound.
The reheat temperature  is then  given by
\be \label{eq4}
 T_{RH} \simeq  1.2 g_{_{RH}}^{-1/4}\sqrt{M_P \Gamma_{\varphi} } \simeq  
 1 \left(0.1 \over B_a \right)^{1/2} \left( 10^{12} \GeV \over F_a \right) 
  \left( M_{\varphi} \over 300 \GeV \right)^{3/2} \GeV \,,
\ee
where $g_{_{RH}}\equiv g_*(T_{_{RH}})$ 
counts the effective number of relativistic
degree of freedom at $T_{RH}$.
The entropy production factor
$S_{\rm after}/S_{\rm before}$ for this reheating is
of order $V_0/m_{3/2}^3 T_{RH}$ which is roughly of order
$10^2(M_P/m_{3/2})^{(5n-1)/(2n+2)}$.
This huge entropy dumping at  relatively late time was 
considered as a promising source for erasing out various 
unwanted cosmological  relics, especially,  cosmologically
dangerous string moduli \cite{ls2}.

In order to evaluate $(\rho_a)_{NS}$, one needs to know
whether axions produced by the late flaton decay 
have ever been in thermal equilibrium with 
the thermalized plasma of normal light particles.
If axions were in thermal equilibrium at some moment
but later frozen out  at  temperature $T_f$,
we would have 
\be
\left({\rho_a \over \rho_{\nu}}\right)_{NS}=
\frac{4}{7}\left(\frac{43/4}{g_*(T_f)}\right)^{4/3}.
\ee
However if axions have never been in equilibrium,
$(\rho_a)_{NS}$ is simply determined by
the {\it effective} branching ratio  $B_a$
measuring how large fraction of flatons are 
converted into axions during the reheating. Roughly 
$B_a \simeq \Gamma_{a}/\Gamma_{\rm tot}$
with the decay width  $\Gamma_a$  
of $\varphi\rightarrow 2a$, however axions can be
produced also by
the secondary decays of
the decay products of flatons.
For unthermalized axions at $T_{RH}$, the ratio between 
$\rho_a$ and the energy
density  $\rho_r$ of thermalized radiation 
would be simply $B_a/(1-B_a)$.
We then have
\be
\left({\rho_{a} \over \rho_{\nu}}\right)_{NS}=
\frac{43}{4}\frac{B_a}{1-B_a}\frac{4}{7}\left(43/4\over
g_{_{RH}} \right)^{1/3}.
\ee

In order to see whether axions have ever been in thermal
equilibrium, let us consider the axion interaction rate
$\Gamma_{\rm int}=\lag \sigma v\rag N_r$
where $\sigma$ denotes the cross section for the
axion scattering off the thermalized  radiation with
energy density $\rho_r$ and number density $N_r$.
A careful look at  of the reheating process  indicates that
$\rho_r\sim R^{-3/2}$, $N_r\sim R^{-1/2}$, and
$\rho_{\varphi}\sim R^{-3}e^{-t\Gamma_{\rm tot}}$
during the reheating period
between $t_0$ and
$t_{_D}\simeq \Gamma_{\rm tot}^{-1}$ where $R$ denotes the scale
factor and $t_0$ corresponds to the time
when the relativistic particles produced by
the flaton decay become the major part of the radiation \cite{dfst}.
A simple dimensional analysis implies that
the axion cross section 
can be written as
$\sigma=(\gamma_1+\gamma_2 (m/E)^2)/4\pi F_a^2$,
where $E$ denotes the center of mass energy,
$\gamma_{1,2}$ are dimensionless constants of order unity
or less, and $m$ corresponds to 
 the mass of target particle.
We then have $\lag \sigma v\rag = 
(\gamma_1+\gamma_2 (m/\lag E_0\rag)^2(R/R_0)^2)/4\pi F_a^2$.
With the  informations given above, it is straightforward 
to see that the ratio $Z=\Gamma_{\rm int}/H$ is an increasing
function of $R$ during the reheating period of $t_0<t<t_{_D}$.

Let us now consider the behavior of $Z=\Gamma_{\rm int}/H$
after the reheating. 
At $t>t_{_D}$ with $T< T_{RH}$,
the entropy production almost ends and thus
$N_{r}\sim R^{-3}\sim T^3$, $H\sim R^{-2}\sim T^2$  as in the standard
radiation dominated universe with an adiabatic expansion.
Using Eq.~(3) with $M_{\varphi}\simeq m_{3/2}$ and Eq.~(5), we  find
\bea
{\Gamma_{\rm int} \over H} &=& 3\times 10^{-2}\bar{\gamma}
 g_*^{1/2} 
\frac{T M_P}{F_a}^2 \nonumber \\
&=& 10^{-2} \bar{\gamma}\left({T\over T_{RH}}\right)
\left(\frac{g_*}{10^2}\right)
\left(\frac{M_{\varphi}}{M_P}\right)^{3(n-1)/2(n+1)},
\eea
where $\bar{\gamma}=(\gamma_1+\gamma_2 (m/\lag E\rag)^2)$.

For $n>1$, the above result for $t>t_{_D}$ together with the fact that
$Z=\Gamma_{\rm int}/H$ is an increasing function of $R$ during $t_0<t<t_{_D}$
readily implies that $Z\ll 1$ and thus axions have never been
in equilibrium.
For the case of $n=1$, we need a bit more discussion
about the size of $\bar{\gamma}$.
Obviously at tree level, any
nontrivial axion couplings to $SU(2)\times U(1)$ {\it non-singlet}
fields  
arise as a consequence of $SU(2)\times U(1)$ breaking.
In other words, tree level axion couplings to normal fields
are induced by
the mixing with the Higgs
doublets. As a result, tree level axion
couplings can be described effectively by  dimensionless coupling
constants which are of order $m/F_a$ where $m$ corresponds
to the mass
of the particle that couples to the axion.
This means 
that the energy dependent part of the axion cross section,
i.e. the $\gamma_2$-part, is  due to
tree level axion couplings, while the energy independent
$\gamma_1$-part is due to the loop-induced axion couplings
like $\frac{\alpha_s}{4\pi F_a}aG_{\mu\nu}\tilde{G}^{\mu\nu}$.
As a result, $\bar{\gamma}$ is suppressed
either by the loop factor $(\frac{1}{8\pi^2})^2$
or by  the relativistic factor $(m/\lag E\rag)^2$.
Then we can 
safely take $\bar{\gamma}\ler 1$, implying $Z\ll 1$ for the case 
of $n=1$ also.

In the above, we have argued that $Z=\Gamma_{\rm int}/H\ll 1$
and thus axions have never been in thermal equilibrium.
Then the axion energy density at nucleosynthesis is given
by Eq.~(7) and the NS bound (4) leads to
\be \label{bound}
\frac{B_a}{1-B_a} \leq 0.24 \, \left( \delta N_{\nu}\over 1.5 \right) 
             \left(g_{_{RH}}\over 43/4\right)^{1/3}.
\ee
The above nucleosynthesis limit on $B_a$ depends mildly upon the reheat
temperature $T_{RH}$ through the factor 
$(g_{_{RH}}/g_{_{NS}})^{1/3}$ where $g_{_{NS}}\equiv g_*(T_{_{NS}})=43/4$,
while it is rather sensitive to
the discordant number $\delta N_{\nu}$ which is presumed here
to be in the range $0.1 \sim 1.5$
\cite{nsold,nsnew}.
For $T_{RH}$ above 0.2 GeV but below the superparticle mass, 
we have $g_{_{RH}}/g_{_{NS}}=
6 \sim 10$,  while  $g_{_{RH}}/g_{_{NS}}=1\sim 3$ for $T_{RH}< 0.2$ GeV.
We thus have just 
a factor two variation of the limit  when  $T_{RH}$ varies 
from the lowest allowed value 6 MeV \cite{lps} to the superparticle 
mass of order 100 GeV.
In summary,  the NS limit (9) indicates that 
we  need to tune  the effective branching ratio
$B_a$ to be less than $1/3 \sim 0.02$
for $\delta N_{\nu}=0.1 \sim 1.5$.
\bigskip

{\em Implication on flaton couplings. }
We now discuss  the implications of
the NS bound (9) for generic supersymmetric
axion models with a radiative mechanism generating
the axion scale. 
Since the models under consideration involve
too many unknown free parameters, we just examine 
how plausible it is  that the NS limit (9) is satisfied 
for  the unknown parameters 
simply taking  their natural values.
To proceed, let us write the flaton
couplings responsible for the flaton decay
as
\be
{\cal L}_{\varphi}={\cal L}_{PQ}+{\cal L}_{
{SSM}},
\ee
where ${\cal L}_{PQ}$ describes the couplings
to the  fields in  PQ sector, while ${\cal L}_{
{SSM}}$ describes  the couplings to the fields in supersymmetric
standard model (SSM) sector.
Schematically ${\cal L}_{{PQ}}$ is given by
\be
{\cal L}_{{PQ}}=
\frac{\varphi}{2F_a}\left(M_{\varphi}^2a^2+M_{\varphi}^2\varphi^{\prime 2}
+(M_{\tilde{\varphi}}\tilde{\varphi}\tilde{\varphi}+{\rm h.c})\right)
\ee
where $a$, $\varphi^{\prime}$, and $\tilde{\varphi}$ denote
the axion, other flaton, and flatino respectively.
In the above, we have ignored the model-dependent
dimensionless coefficients
of each terms which are of order unity in general.

The flaton couplings to SSM fields are more  model dependent. 
In DFSZ type models,
flaton couplings to the  SSM sector
are essentially due to
the mixing 
with the Higgs doublets.
Then flaton couplings can be read off 
by making the replacement
\be
H_i \to  v_i+ {x_{ij}v_j \over F_a} \varphi,
\ee 
where $v_i=\lag H_i\rag$,  $x_{ij}$'s are model-dependent
coefficients which are generically of order unity.
Then again schematically
\be \label{s-two}
 {\cal L}_{SSM} = {\varphi \over F_a} \left\{ 
(M_1 \chi\chi^{\prime} 
+M_2 \chi \lambda +{\rm h.c.})
   +M_3^2 A_{\mu}A^{\mu}+ M_4^2zz^{\prime}+M_5^2 |z|^2 \right\},
\ee
where
$z$ and $z^{\prime}$
denote  spin zero fields in the 
SSM, e.g. 
squarks, sleptons and Higgs, with their fermionic
partners $\chi$ and $\chi^{\prime}$,
while $(A^{\mu}, \lambda)$ stands for  the   gauge multiplets
which become massive due to the Higgs doublets VEVs,
i.e $W$ and $Z$.
The order of magnitude estimate of the dimensionful coefficients
leads to:
$M_1\simeq M_{\chi}$, $M_2\simeq M_3\simeq  M_W$, $M_4^2\simeq M_{\chi}(
A+\mu \cot\beta)$, and $M_5^2\simeq M_{\chi}^2+M_W^2\cos 2\beta$,
where $M_{\chi}$ and $M_W$ denote the masses of 
$\chi$ and $W$ respectively, $\tan \beta=v_2/v_1$,
and again we have ignored
the coefficients of order unity.

As is well known, besides DFSZ type models,
there are another interesting class of
axion models named as hadronic axion models.
In hadronic axion models, all SSM fields carry 
{\it vanishing} PQ charge and as a result
flaton couplings to SSM fields appear as  loop effects.
As an example of  supersymmetric hadronic axion model
with a mechanism generating the axion scale radiatively,
let us consider a model with
\be
W=k\frac{\phi_1^{n+2}\phi_2}{M_P^n}+h_QQQ^c\phi_1+\cdots,
\ee
where $\phi_{1,2}$ are gauge singlet flatons,
and $Q$ and $Q^c$ stand for additional heavy quark and antiquark
superfields.
Again the soft mass-squared  of  $\phi_1$
becomes {\it negative} at scales around $F_a$
by the radiative corrections involving the
strong Yukawa coupling $h_QQQ^c\phi_1$, thereby 
generating 
the axion scale as
$\langle \phi_{1}\rangle\simeq \langle \phi_2\rangle\simeq F_a$.
A peculiar feature of this type 
of hadronic axion models is that at tree level flatons
do not couple to SSM fields,
while there
are nonzero couplings to PQ fields as Eq.~(11).
Flaton couplings to SSM fields are then induced by
the loops of $Q$ and $Q^c$, yielding
\be
{\cal L}_{SSM}=\frac{\alpha_s}{2\pi}\frac{\varphi}{F_a}\left(
\frac{1}{4}G^a_{\mu\nu}G^{a\mu\nu}
+i\bar{\lambda}^a\gamma\cdot\partial\lambda^a)\right),
\ee
where $(G^a_{\mu\nu},\lambda^a)$ denotes the gluon supermultiplet 
(and possibly other gauge multiplets)
and again we ignored dimensionless coefficients of order unity.

It is now easy to notice that, due to the loop suppression
in ${\cal L}_{SSM}$, most of 
oscillating flatons in  hadronic axion
models decay first into either axion pairs, or lighter flaton pairs, or 
flatino pairs, as long as the decays are
kinematically allowed. Lighter flatons would experience similar
decay modes, while flatinos decay into axion plus a lighter flatino.
Then in the first round of reheating,
flatons are converted into either axions
or the lightest flatinos. The lightest flatinos will eventually
decay into SSM particles. Because of
kinematical reasons, e.g. the mass relation 
$M_{\varphi}> 2 M_{\tilde{\varphi}}$  and the phase space suppression
factor  $(1-4M_{\tilde{\varphi}}^2/M_{\varphi}^2)^{1/2}$
in the decay $\varphi\rightarrow 2\tilde{\varphi}$, 
more than half of the original flatons
would be converted into axions, i.e. the effective branching ratio
$B_a\ger 1/2$,  {\it unless} 
the flaton coupling to the  {\it lightest} flatino is unusually large.
This is in conflict with the NS limit (9) even for
the most conservative choice $\delta N_{\nu}=1.5$,
implying that hadronic axion models with a radiative mechanism
can be compatible with the big-bang nucleosynthesis
only when the models  are tuned to 
have an unusually large flaton coupling to the lightest flatino.

In  DFSZ type models, flatons have  tree level couplings to SSM
fields which are of order $M_{_{\rm SSM}}/F_a$
or  $M_{_{\rm SSM}}^2/F_a$
where $M_{_{\rm SSM}}$ collectively denotes 
the mass parameters in the SSM, e.g. $M_t$, $M_W$, $\mu$, $A$,
and so on [see Eq.~(13) and the discussions below it]. 
Thus if $M_{\varphi}\gg M_{_{\rm SSM}}$,
the reheating procedure would be similar as 
that of hadronic axion models and then the NS limits provides
a meaningful constraint on the flaton couplings to the PQ sector.
For the case that $M_{\varphi}$ is comparable to $M_{_{\rm SSM}}$,
the most conservative choice 
of $\delta N_{\nu}=1.5$ would  not provide any meaningful
restriction on DFSZ type models.
However 
it is still  nontrivial
to achieve $B_a$ significantly smaller than 1/10.
A careful examination of the flaton couplings in DFSZ
type models
suggests  that,  among the decays into SSM particles,
the decay channels to the top ($t$) and/or stop ($\tilde{t}$)
 pairs are most important.
Flaton coupling to the top
(stop)   is of order $M_t/F_a$  
($M_{\tilde{t}}^2/F_a$), while the coupling to the axion is of order
$M_{\varphi}^2/F_a$. As a result,  $B_a$ significantly smaller 
than $1/10$ implies that the flaton couplings to the top and/or
stop are unusually large in view of
the relation 
$M_{\varphi}> 2M_t$ ($2M_{\tilde{t}}$).
One of the efficient way to achieve such a small $B_a$ is 
to assume that  there is a sort of  mass hierarchy  
between the lighter stop mass-squared  $M^2_{\tilde{t}_1}$
and the heavier stop-mass squared $M^2_{\tilde{t}_2}$,
allowing  for instance  
$4M^2_{\tilde{t}_1} <M^2_{\varphi}
< \frac{1}{4}M_{\tilde{t}_2}^2$.
This would be the case when
$M^2_{\tilde{t}}+M_t^2 \simeq M^2_{\tilde{t}^c}+M_t^2 \simeq
M_t(A+\mu\cot\beta)$ where $M^2_{\tilde{t}}$ and $M^2_{\tilde{t}^c}$
denote the soft  squark masses.
Since the flaton couplings to stops are determined not
only by the mass parameters  (e.g. $M_t$ and $A$)
but also by additional dimensionless
parameters $x_{ij}$ defined in Eq.~(12),
the flaton coupling to the lighter stop $\tilde{t}_1$
would be  of order $M^2_{\tilde{t}_2}/F_a$,
not the order of $M^2_{\tilde{t}_1}/F_a$.
To be more explicit, let us write this coupling as
$x_1M^2_{\tilde{t}_2}\varphi |\tilde{t_1}|^2/F_a$.
With the flaton-axion coupling given by 
$x_a M^2_{\varphi} \varphi a^2/2F_a$, we find
\be
{\Gamma_a \over \Gamma_{\tilde{t}_1}}={1 \over 32} \left({x_a
\over x_1}\right)^2\left(\frac{2M_{\varphi}}{M_{\tilde{t}_2}}\right)^4
\left(1- \frac{4M_{\tilde{t}_1}^2}{M_{\varphi}^2}\right)^{-1/2}.
\ee
This shows that  $B_a$ can be smaller than about $10^{-2}$
for the parameter range: $x_a\approx x_1$ and 
$4 M^2_{\tilde{t}_1} <M^2_{\varphi}
< \frac{1}{4}M^2_{\tilde{t}_2}$.
\bigskip

{\em Relaxation of the bound on $F_a$. }
The reheat temperature 
can not be arbitrarily low in order to be
compatible with the big bang 
nucleosynthesis.
Since flatons
produce  large number of hardrons, the bound $T_{RH} > 6$ MeV has to be
B
met \cite{lsss}.  With Eq.~(5), this leads to the upper bound:
\be \label{new}
 F_a \ler 2 \times 10^{14} 
           \left(0.1 \over B_a \right)^{1/2} 
\left(M_\varphi \over 300 \GeV \right)^{3/2} \GeV \,.
\ee
Once  one uses the relation $F_a\simeq (M_{\varphi}M_P^n)^{1/n+1}$,
this means that only $n=1$, 2, and 3 are allowed by the big-bang
nucleosynthesis.

As is well known, another upper bound on the axion scale can be
derived by requiring that the 
coherent axion energy density produced by an initial
misalignment 
should not exceed the critical density \cite{pww}.
If there is no entropy production after the axion start
to oscillate at around
$T\simeq 1$ GeV, this lead to the usual bound:
$F_a\ler 10^{12}$ GeV.
When $n=2$ or 3, 
the corresponding axion
scale $F_a\simeq (M_{\varphi}M_P^n)^{1/n+1}$ would exceed this bound.
However in this case,
the reheat temperature \refs{eq4} goes  
below 1 GeV.
Then the coherent axions may be significantly diluted by 
the entropy dumped from flaton decays, thereby 
allowing  $F_a$ much bigger than $10^{12}$ GeV \cite{dfst}.

Axion production in matter-dominated universe, e.g. flaton oscillation
dominated universe,  has been considered in 
Ref.~\cite{lsss,lps} assuming $m_a(T) \propto T^{-4}$.   
For our computation,  we take the power-law fit 
of the temperature dependent axion mass \cite{tur}:
$$m_a(T) \simeq 7.7\times10^{-2}\, m_a(T\!=\!0)(\Lambda_{QCD}/T)^{3.7}\,.$$ 
Axion oscillation starts at $T_a$ for which $m_a(T_a)=3H(T_a)$: 
\be\label{Ta}
 T_a \simeq 0.9 
  \left(\Lambda_{QCD} \over 200 \MeV \right)^{0.48}
  \left(M_{\varphi} \over 300 \GeV \right)^{0.39}
  \left(10^{12} \GeV \over F_a \right)^{0.39} \GeV \,.
\ee
We refer the reader to paper \cite{st} for the available  formulae.
If $T_a>T_{RH}$, the coherent axion energy density is diluted by the entropy
produced between $T_a$ and $T_{RH}$.  At the end of the entropy
dumping (around $T_{RH}$), the coherent axion number density in unit
of the entropy density is given by
$Y_f \simeq {\theta^2F_a^2 m_a(T_a) R_a^3 /S_f }$
where $\theta$ denotes the  initial  misalignment angle of
the axion  field,
$R_a$ is the scale factor at $T_a$ and $S_f$ is the total entropy at
$T_{RH}$.  The ratio of the
axion energy density 
to the critical energy density at present is  given by
\bea
\Omega_a h_{50}^2&\simeq&  
3.3\times 10^{17} \left(\frac{F_a}{10^{12} \, {\rm GeV}}\right)^{1.5}
\left(\frac{\Gamma_{\varphi}}{\rm GeV}\right)^{0.98} 
\left(\Lambda_{QCD}\over 200 \, \MeV \right)^{-1.9}\nonumber \\
&\simeq & 1 \left(0.1 \over B_a\right)
\left( 10^{12} \GeV \over F_a \right)^{0.44}
\left(\frac{M_{\varphi}}{300 \, {\rm GeV}}\right)^{2.9}
\left( \Lambda_{QCD} \over 200 \, \MeV \right)^{-1.9}
\eea
where we have used 
$\Gamma_{\varphi}\simeq B_a^{-1}M_{\varphi}^3/32\pi F_a^2$.
The above result  is valid only for $n\geq 2$ yielding $T_{RH}<T_a$.
As we have anticipated, it shows that the case of $n=2$ or 3
with $F_a\simeq (M_{\varphi}M_P^n)^{1/n+1}$
yields  a coherent axion
energy density not exceeding the critical density although
the corresponding $F_a$ exceeds $10^{12}$ GeV.
Furthermore, in this case of $n=2$ or 3,
{\it axions can be a good dark matter
candidate} for an appropriate value of $M_{\varphi}$, which
was  not possible for $n=1$.

We remark that diluting the coherent axions
with $T_{RH}<T_a$ is allowed 
only when  R-parity is broken.
If not,
stable lightest supersymmetric particles (LSP)
produced after the flaton decay would overclose 
the universe.  This can be avoided if the reheat temperature is
bigger than the decoupling temperature of LSP which is typically 
$M_{LSP}/20$.  However this is usually above 1 GeV, i.e.
above $T_a$.
Consequently, the usual upper
bound $F_a \ler 10^{12}$ GeV can not be relaxed when the reheat temperature
is bigger than $M_{LSP}/20$.
We stress here that
even when R-parity is broken and thus
LSP cannot be a dark matter candidate,
coherent axions can be a viable dark matter candidate when $n=2$ or 3.
\bigskip

{\em Baryogenesis. }
Thermal inflation driven by PQ flatons may dilute away any
pre-existing baryon asymmetry.  However, PQ flatons themselves  can 
produce baryon asymmetry after the reheating through the
DH mechanism \cite{dh}.  A complicated Affleck-Dine type baryogenesis
after thermal inflation has also been explored in Ref.~\cite{sky}.  
Our previous discussion of flaton couplings in DFSZ type
models indicates that flatons  going to
stops can be the  most efficient decay channel.  
The decay-produced stops subsequently decay to generate a baryon asymmetry 
provided that the baryon-number violating operator, e.g., $\lambda''_{332}
U^c_3 D^c_3 D^c_2$ and the corresponding complex trilinear soft-term
are present.  Note that the PQ symmetry [see Eq.~\refs{supo}]
can be arranged so that
dangerous lepton-number violating operators $LQD^c, LLE^c$ are 
forbidden for the proton stability.

In order for the baryon asymmetry  not to be erased
the reheat temperature \refs{eq4} has again to be less than few GeV
\cite{dh}. This again means that the DH mechanism
can work only for $n=2$ or 3 [see Eqs.~(3) and (5)].
The produced baryon asymmetry is 
\be \label{ba1}
 \eta \equiv {n_B \over n_\gamma} 
      \simeq 5.3 {T_{RH} \over M_\varphi} \Delta B\,,
\ee
where $\Delta B$ is the baryon asymmetry generated by each flaton
decay into stop-antistop pair.
Using  Eq.~\refs{eq4} and the estimate of  $\Delta B$ given
in \cite{dh}, we find 
\be \label{ba2}
 {\eta \over 3\times10^{-10}} \simeq 
|\lambda''_{332}|^2
\left({\rm arg}(Am_{1/2}^*)\over 10^{-2}\right)
\left(0.1 \over B_a \right)^{1/2}
\left(10^{14} \GeV \over F_a \right)
 \left( M_\varphi \over 300 \GeV \right)^{1/2},
\ee
where  ${\rm arg}(A m_{1/2}^*)$ denotes the CP violating
relative phase which is constrained to be less than
$10^{-2}$ for superparticle masses of order 100
GeV \cite{edm}.
For $n=3$, the desired amount of baryon asymmetry
can be achieved only when  $\lambda''_{332}$
is of order unity, while for $n=2$ it can
be done  with a smaller $\lambda''_{332}$.
\bigskip

In conclusion, we have examined  
some  cosmological consequences of supersymmetric
axion models in which the axion
scale is radiatively generated as
$F_a\simeq (m_{3/2} M_P^n)^{1/n+1}$.
In such models, the early universe inevitably experiences 
a period dominated by the coherent oscillation
of  PQ flatons which start to oscillate
at  temperature around $m_{3/2}$. 
Then a significant amount of oscillating PQ flatons can decay
into axions, thereby yielding 
a too large axion energy density at the time
nucleosynthesis.
This consideration puts a limit on the effective branching
ratio $B_a$ measuring how large fraction
of oscillating flatons are converted into axions: it should
be less than $1/3 \sim 0.02$ depending upon our choice
of the allowed extra number of neutrino species $\delta N_{\nu}=
0.1 \sim 1.5$.  
Models of hadronic axion with a radiative mechanism
would yield  $B_a\ger 0.5$ unless the flaton coupling
to the lightest flatino is unusually large.
This is essentially because
the flaton couplings to SSM are loop suppressed compared
to the couplings to PQ sector.
DFSZ type models with a radiative
mechanism is more interesting since it can  provide
a rationale for the size of the  $\mu$  term (and also
the scale for neutrino masses).
If the flaton mass $M_{\varphi}\gg M_{_{\rm SSM}}$ denoting
the typical mass in supersymmetric standard model,
DFSZ type models would also  suffer from the same difficulty 
as that of hadronic axion models.
However for $M_{\varphi}$ comparable to $M_{_{\rm SSM}}$,
requiring $B_a$ to be about 1/10
does not provide any meaningful constraint on DFSZ type models.
If one wishes to achieve a smaller  $B_a$,
say about $10^{-2}$ in DFSZ type models,  one then needs a kind of
tuning of the model.
Flaton decays into the lighter  stops  is then picked out as
one of the 
efficient decay  channels leading to such a small $B_a$
provided 
$4M^2_{\tilde{t_1}}< M^2_{\varphi} < \frac{1}{4}M^2_{\tilde{t_2}}$.

Another  interesting cosmological consequence
of decaying flatons is  the relaxation of the cosmological
upper bound on  the axion scale.
For the axion scale bigger than $10^{12}$
GeV, the entropy production by PQ flatons ends 
after  the axion field  starts
to oscillate by QCD instanton effects,
thereby diluting the
coherent axion energy density in a rather natural
way.  With this late time entropy production 
by PQ flatons, the upper bound on the axion scale $F_a$
can be pushed up to about  $10^{15}$ GeV, but at the expense
of breaking R-parity to avoid a too large mass density
of relic LSP.
Then the integer $n$ which determines the axion scale
in terms of $m_{3/2}$ and $M_P$  can take
$n=1$, 2 or 3. 

It is likely that any pre-existing baryon 
asymmetry  is completely diluted by the huge entropy dumping
in  thermal inflation scenario.
As the PQ flatons   are expected to decay dominantly into stops, 
the DH mechanism for the late time  baryogenesis  
can work in a natural manner when $n=2$ or 3 so that the 
reheat temperature does not exceed  1 GeV.
With broken  R-parity, LSP is no more stable
and can not be a dark matter candidate.
In this scenario,
coherent axions can  provide a critical mass density of the universe
by saturating the cosmological bound on $F_a$ which  now can be
as large as $10^{15}$ GeV.

Interestingly enough,  
we now observe that the case of $n=2$ or $n=3$  provides a very 
concordant cosmological scenario:
(i) a proper baryon asymmetry is generated 
by the DH mechanism using baryon-number
violating interaction $\lambda'' U^cD^cD^c$,
(ii) potentially dangerous
coherent axions (with $F_a \gg 10^{12}$ GeV) are diluted by
the late time entropy production,
(iii) both  the baryogenesis and  axion dilution
require R-parity to be broken, and then diluted
coherent axions constitute dark matter in the universe,
\bigskip

{\bf Acknowledgments}:
This work is supported in part by KOSEF through CTP of
Seoul National University (KC, JEK), KAIST Basic Science Research
Program (KC), KOSEF Grant 951-0207-002-2 (KC), Basic Science
Research Institutes Program, BSRI-96-2418 (JEK), BSRI-96-2434 (KC),
and the SNU-Nagoya Collaboration Program of Korea Research Foundation
(KC, EJC, JEK).  EJC is a Brain-Pool fellow.

\end{document}